\documentclass[reprint, twocolumn]{revtex4-1}

\usepackage{amsmath}    
\usepackage{graphicx}   
\usepackage{verbatim}   
\usepackage{color}      
\usepackage{subfigure}  
\usepackage{hyperref}   
\begin{document}

\title{Reconfigurable random bit storage using polymer-dispersed liquid crystal}

\author{Roarke Horstmeyer*}
\affiliation{California Institute of Technology, Department of Electrical Engineering, Pasadena, CA 91125}
\author{Sid Assawaworrarit*}
\affiliation{California Institute of Technology, Department of Electrical Engineering, Pasadena, CA 91125}
\author{Changhuei Yang}
\affiliation{California Institute of Technology, Department of Electrical Engineering, Pasadena, CA 91125}


\begin{abstract}
We present an optical method of storing random cryptographic keys, at high densities, within an electronically reconfigurable volume of polymer-dispersed liquid crystal (PDLC) film. We demonstrate how temporary application of a voltage above PDLC's saturation threshold can completely randomize (i.e., decorrelate) its optical scattering potential in less than a second. A unique optical setup is built around this resettable PDLC film to non-electronically save many random cryptographic bits, with minimal error, over a period of one day. These random bits, stored at an unprecedented density (10 Gb/mm$^{3}$), can then be erased and transformed into a new random key space in less than one second. Cryptographic applications of such a volumetric memory device include use as a crypto-currency wallet and as a source of resettable ``fingerprints" for time-sensitive authentication. (*)Authors contributed equally to this work.
\end{abstract}

\maketitle

A common problem facing all cryptographic systems is how to secure their secret keys from malicious attack. An effective key storage medium should be both challenging for someone to copy or tamper with, as well as quickly erasable (i.e., easily destroyed when circumstances require). Many cryptosystems, both historic and current, fail to achieve these two properties simultaneously. For example, the Enigma machine was complex enough to prevent duplication for several years during WWII, but could not be easily destroyed. At the same time, it was common to distribute one-time cipher pads printed on flammable paper. The pads could be quickly burned, but were easily copied and distributed when recovered by an enemy. 

Today, digital electronic memory stores nearly all of our cryptographic random keys. Unfortunately, digital storage neither prevents copying nor is easily erasable, even if designed towards such ends. Many cryptographic systems store random keys in non-volatile memory (e.g., EEPROMÕs), for which several well-known invasive attacks exist~\cite{Anderson:98}. Semi-invasive attacks have also copied the sensitive bits stored in volatile forms of memory, like SRAM~\cite{Skorobogatov:03} and DRAM~\cite{Halderman:08}. Even systems designed to ensure that sensitive information disappears from transient memory when powered off have been easily circumvented~\cite{Halderman:08,Samyde:02}. Apart from external attack, an adversary may internally embed untrustworthy pseudorandom number generation software~\cite{Shumow:07} or electronic hardware~\cite{Karri:10,Becker:13} without revealing any breach of security to the user. 
 
Physical unclonable functions (PUFs) are a recently proposed class of storage device that attempt to increase the difficulty of copying, modeling or probing the contents of digital electronic memory~\cite{Pappu:02}. They use the inherent microscopic physical disorder within a device, often in the form of variations induced during fabrication, to form a unique ``fingerprint" that is extremely challenging to copy or model. Examples include timing offsets in integrated circuits~\cite{Lim:05}, instabilities in volatile memory cells~\cite{Guajardo:06}, capacitances of perturbed films~\cite{Tuyls:06}, and scattering potentials of volumetric materials~\cite{Pappu:02}. Random keys are typically derived from these physical fingerprints after additional digital processing. Most PUFÕs to date remain electronic-based, which unfortunately have a very low key storage density (at most several kilobits per square millimeter) and are challenging to fully reconfigure into a new random state. 

Reconfigurability in modern systems is a desirable property not just for preventing use in the case of device theft. It can also increase the strength of various communication protocols by limiting the amount of encrypted data available for cryptanalysis (i.e., by periodically rotating keys). By ensuring keys disappear in an unrecoverable manner, reconfigurability also enables time-sensitive protocols, e.g. to assist with digital copyright management or to limit building access. Several initial reconfigurable electronic PUF demonstrations have shown limited success~\cite{Kursawe:09, Majzoobi:09,Katzenbeisser:11}, but still do not approach sufficient key-bit densities for many applications of interest. Furthermore, investigations have scrutinized~\cite{Ruhrmair:10} and proven incorrect~\cite{Nedospasov:13} ``unclonability" claims for electronic PUFs  - their two-dimensional surfaces still expose sensitive content to direct measurement or malicious alteration.

Here, we present a reconfigurable optical scattering-based storage (ROSS) mechanism whose volumetric structure is both ``strongly unclonable" (as defined in~\cite{Ruhrmair:10}) and can have its contents quickly erased. To read a fixed number of random bits, we probe a disordered volume of particles with an input coherent optical field (Fig. 1). The distributed particles will scatter light into a unique speckle interference pattern, which a digital detector measures to form into a randomized output. This random output ``key" depends both upon the unique distribution of particles within the scattering volume as well as the shape of the input coherent light field. We can sequentially perturb the input light field's phase using a spatial light modulator (SLM) to output a large number of independent speckle patterns. The large space of possible optical scattering interactions within the particle volume ensures the set of output speckle patterns is effectively uncorrelated, leading to a large set of statistically random keys. The volumetric nature of the scattering structure facilitates very high random bit storage densities: 1 Tb/mm$^{3}$ predicted in theory, and 10 Gb/mm$^{3}$ thus far demonstrated in experiment~\cite{Horstmeyer:13}. 

As first analyzed in \cite{Pappu:02}, the volumetric nature of scattering storage ensures copying, modeling or fully measuring the scattering response of many distributed mesoscopic particles remains a significant challenge. The unclonable security of a stolen ROSS device is limited by the time it would take an attacker to model its complete scattering response, equivalent to characterizing the scattering volume's optical scattering transmission matrix~\cite{Rossum:99}. Several techniques enable such an attack~\cite{Popoff:10} but take upwards of several days per device~\cite{Horstmeyer:13}. A scattering medium with a variable scattering transmission matrix, such as PDLC, may allow a PUF device to periodically reset its stored randomness to extend its effective security lifetime beyond several days. Following, we first experimentally demonstrate how the scattering response of PDLC may be reset, and then use it to construct a device that achieves both our desired security goals of tamper-resistance and erasability, simultaneously.

\begin{figure}[t]
\centering
\includegraphics[scale=0.36]{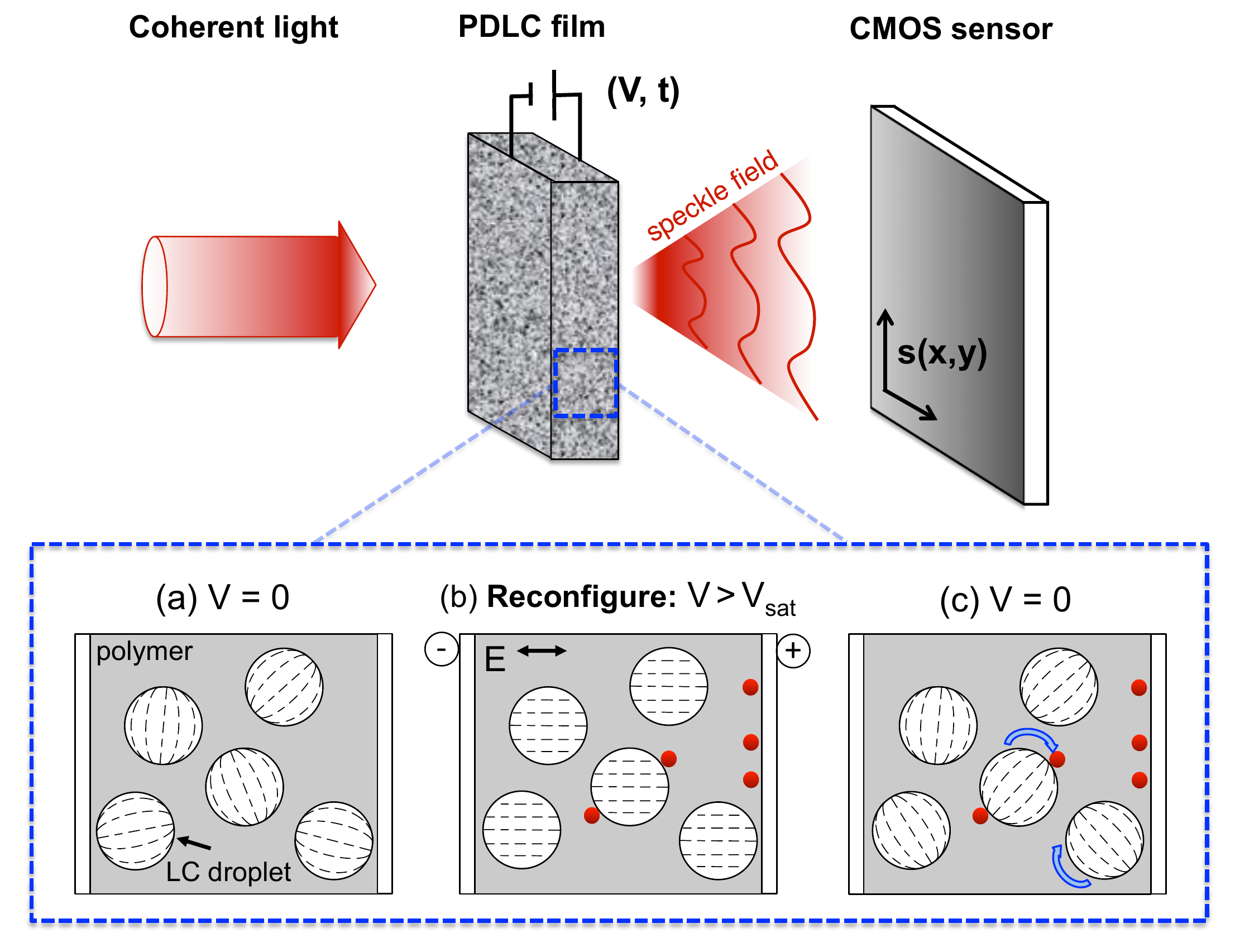}
\caption{The optical scattering response of PDLC film may be reconfigured with temporary application of a large voltage, which introduces mobile ions (red) into the material.}
\label{fig1} 
\end{figure}

PDLC is a well-studied material whose optical transmission properties change with the introduction of a voltage across the film interface. The films employed in this study exhibit a 400 $\mu$m-think optically transparent polymer substrate containing a 20 $\mu$m-thick active layer of sub-micron sized liquid crystal (LC) droplets distributed throughout, sandwiched between a transparent anode and cathode. In the ÒoffÓ state (no voltage), the birefringent LC molecules randomly align themselves between various dislocations (i.e., anchor points) along each droplet's surface (Fig. 1(a)). The boundary of each droplet thus exhibits a random index of refraction mismatch, causing an incident optical field to scatter within the film. The optical response of such a material dense with wavelength-scale particles may be conveniently characterized by a scattering matrix containing complex random Gaussian entries, $T$~\cite{Rossum:99}. In the ``on" state ($V \approx 2$-3V/$\mu$m), the LC's orient themselves along the voltage gradient, aligning the dielectric tensor of all droplets (Fig. 1(b)). The film thus becomes nearly transparent, changing $T$ into an optical transformation that closely resembles the identity matrix. 

When the direct current (DC) voltage $V$ used to keep PDLC in an on-state is above a certain critical saturation value $V_{sat}$, its LC molecules undergo an electrochemical reaction~\cite{Jain:86}. This DC-induced reaction effects both the LCs within each droplet~\cite{Sussman:72} as well as the liquid-polymer and polymer-electrode boundaries, where charge instabilities build up. Specifically, \cite{Perlmutter:96} has shown that the prolonged DC applied to an LC cell introduces mobile ions that selectively adsorb at droplet boundaries. Likewise, \cite{Barbero:90} has derived how free ions at a substrate-LC boundary can shift the LC's anchoring energy, thus rotating its local dielectric tensor. We hypothesize that a combination of the above electrochemical effects cause PDLC film's scattering response to shift after a large DC voltage is applied. Mathematically, we can describe this scattering response shift as a change in a PDLC film's original off-state random scattering matrix $T$ into a new and unpredictable off-state matrix $T'$. As we demonstrate next, the transformation of $T$ into $T'$ effectively ``resets" the space of randomness from which we can derive cryptographic keys. This process cannot be reversed, which gives us our second desired security property of erasability discussed above. 



\begin{figure}
\centering
\includegraphics[scale=0.28]{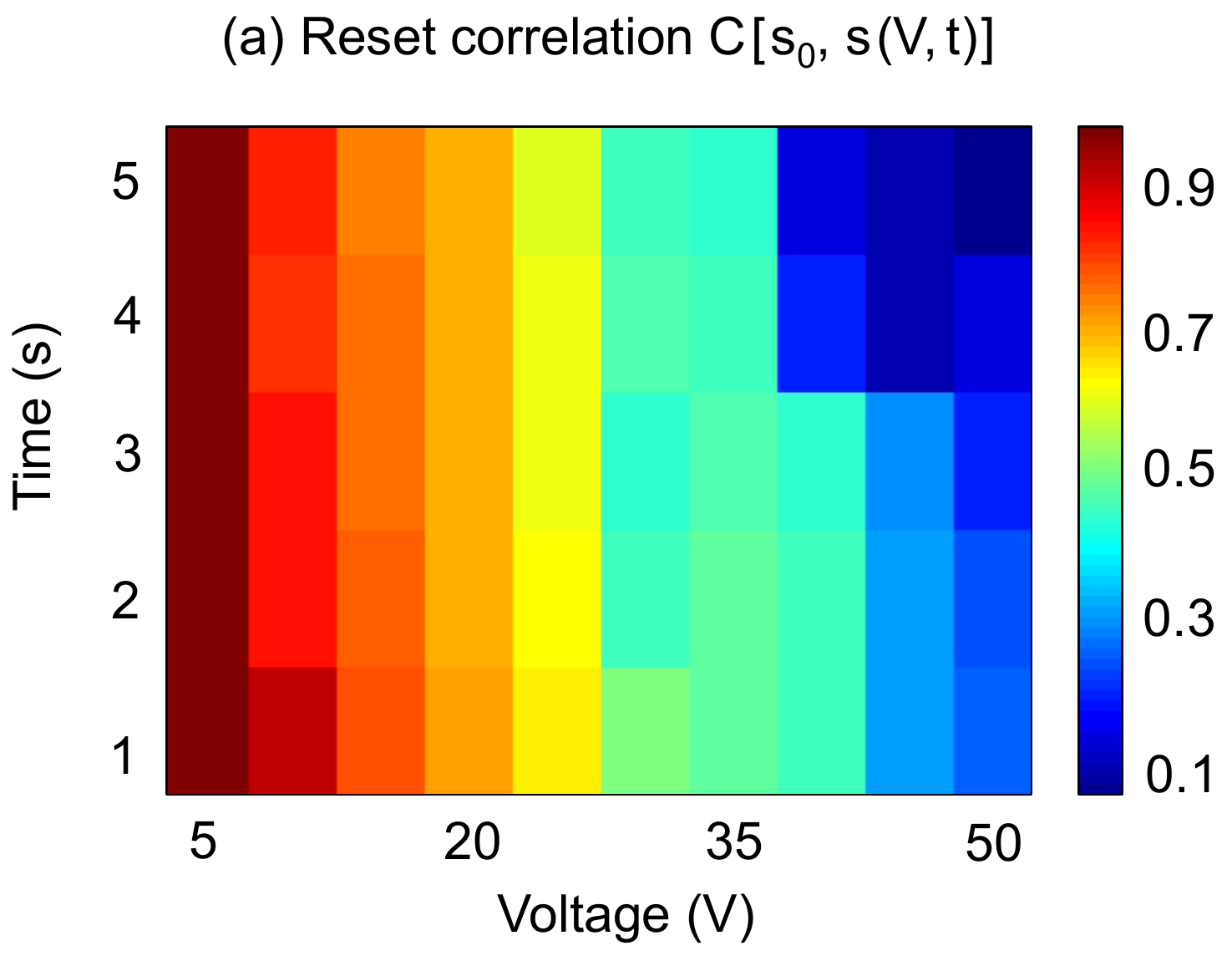}
\includegraphics[scale=0.28]{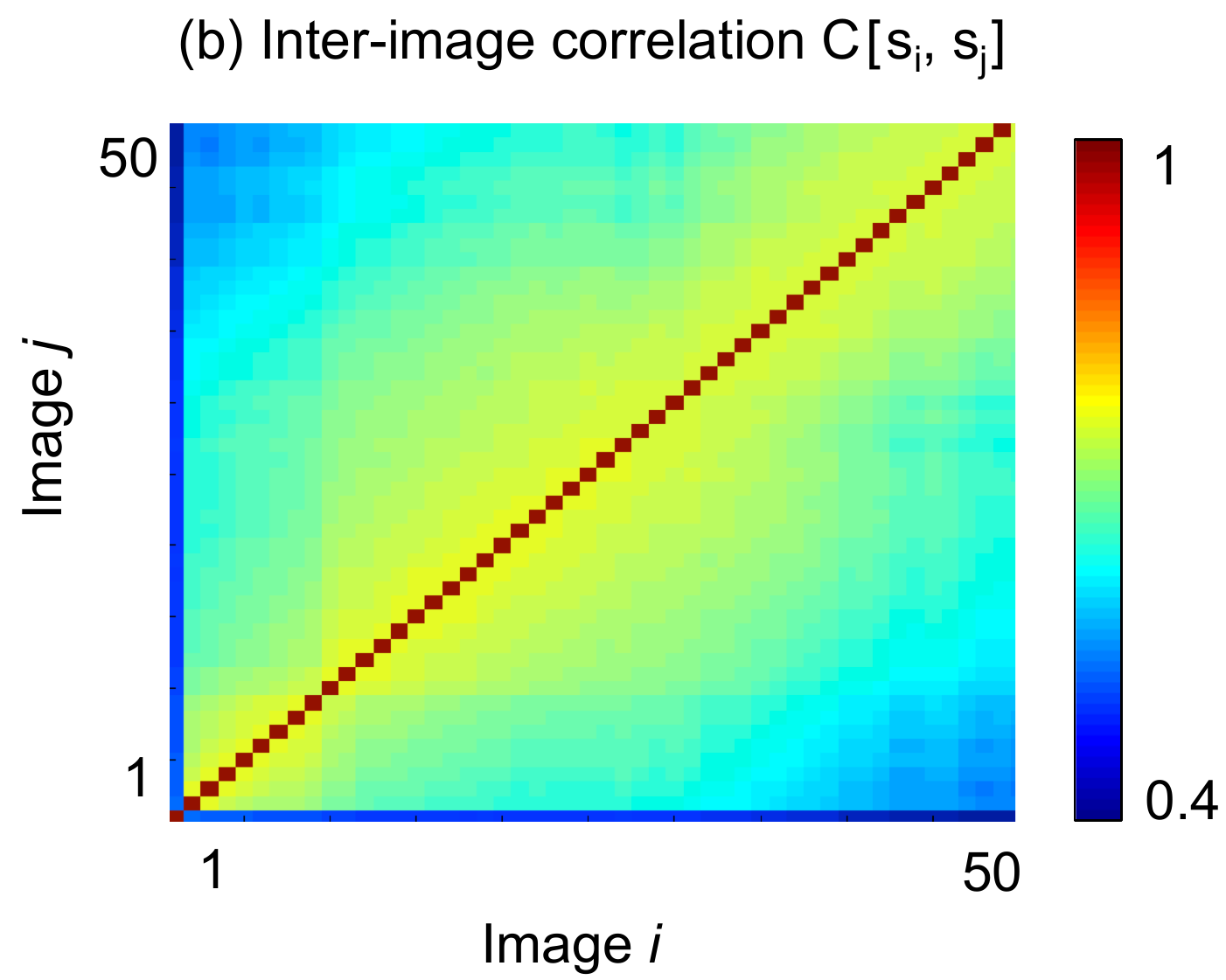}
\caption{Reconfiguring PDLC with an applied voltage. (a) The optical scattering response of PDLC decorrelates to different values as a function of DC voltage $V$ and application time $t$. (b) Cross-correlation $C$ of different speckle images after repeated application of fixed voltage $V=40$ V for duration $t=1$ sec. shows continued film decorrelation.}
\label{fig2} 
\end{figure}

To demonstrate the reconfigurability of PDLC's scattering response, we first illuminate a film with a coherent plane wave of 532 nm light and measure its optical response, $s_{0}(x,y)$, which is the intensity of the speckle field at the digital detector's (Micron CMOS MT9P031) discretized spatial coordinates $(x,y)$ (Fig. 1). Then, we apply a DC voltage $V > V_{sat}$ for a fixed time $t$ across the film surface, during which the film becomes optically transparent. After removing the voltage, we measure a new optical scattering response, $s_{t}(x,y)$ which is significantly different from the original measurement $s_{0}$. We compare $s_{0}$ and $s_{t}$ with a cross-correlation. Performing this experiment for many different values of $V$ and $t$ yields the correlation data in Fig. 2(a), indicating that a 40 V potential applied for several seconds decorrelates the film's scattering response to a minimal value (a new film was used for each measurement to remove any bias). To demonstrate the induced potential continues to produce a random optical response within one film, we repeat this experiment 50 times with the same film fixing $V=40$V and $t=1$ second. All images are significantly (yet not fully) uncorrelated, showing the scattering state does not momentarily leave and return to an original configuration or approach a steady-state molecular configuration, but continues to vary in a semi-random fashion. Improved decorrelation may be achieved by increasing the thickness of the active PDLC material, stacking multiple films along the optical axis, or executing multiple reset operations sequentially over time. 

\begin{figure}
\centering
\includegraphics[scale=0.34]{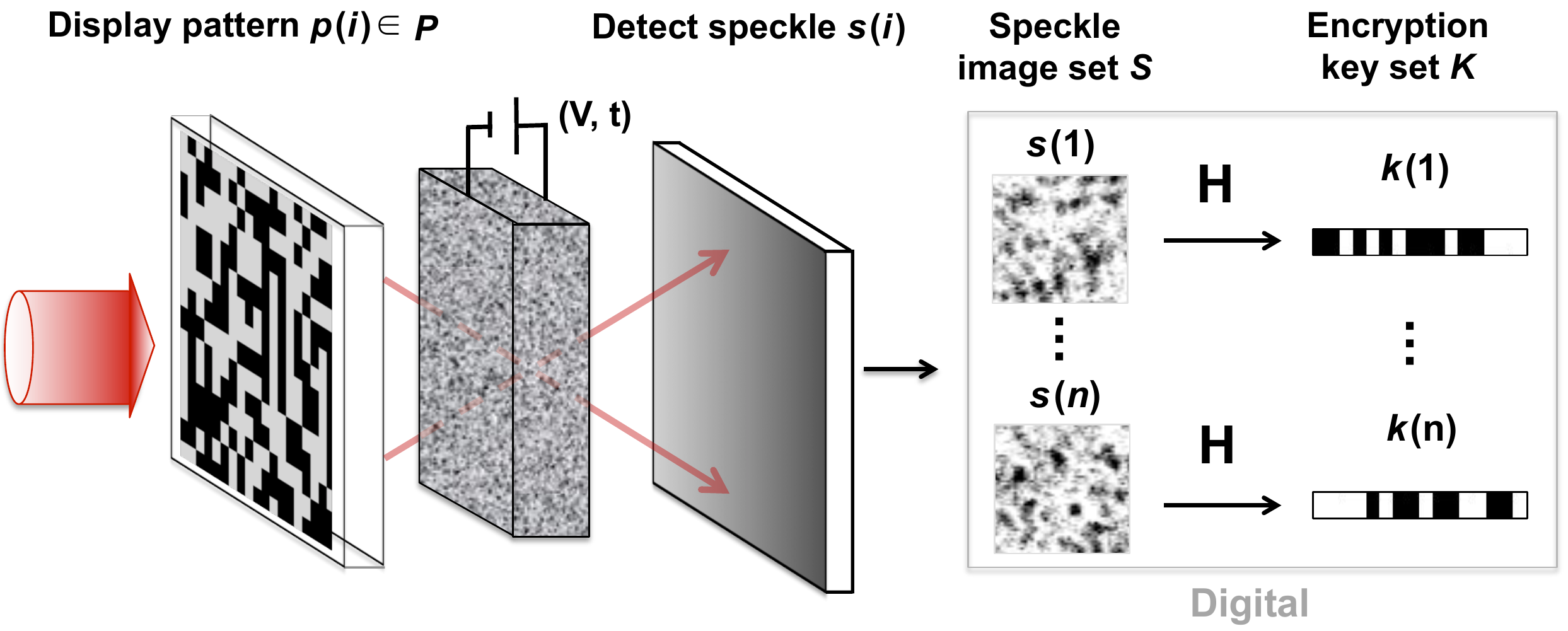}
\caption{The schematic of the ROSS device, where an SLM is used to probe the PDLC material in Fig. 1. Many uncorrelated speckle images $s$ may be detected as many unique binary patterns $p$ are cycled through on the SLM, from which our total key set $K$ is derived.}
\label{fig3} 
\end{figure}

The PDLC volume's random distribution of sub-micron droplets will serve as our ROSS device's random bit ``database". Due to its volumetric structure, physically probing, altering or copying its contexts remains a significant challenge. To selectively address a subset of its stored randomness, we place an SLM (1920 x 1080 pixel Epson HDTV LCD) in front of our PDLC-sensor setup (Fig. 3). These three elements forms our ROSS key storage device. Physically joining the SLM and film with a half-ball lens (radius = 1 cm), and the film to the sensor with a 1 cm quartz cylinder (Mcmaster-Carr 1357T62), helps minimize movement. The SLM controllably varies an ``input" wavefront incident upon the scatterer, to sequentially detect many mutually random speckle intensity pattern ``outputs". Considering our SLM and CMOS array extend along one dimension for simplicity, we may mathematically denote the $i^{th}$ pattern displayed on the SLM as $p(i)$ and the corresponding detected speckle image as $s(i)$. When illuminated with a plane wave, the $i^{th}$ SLM input and detected speckle output are connected by, $s(i)=|Tp(i)|^2$, where $T$ is the volume's unique random scattering matrix. $T$ contains all the inherent randomness that imparts our detected keys with their security. A sufficiently large set of many independent speckle measurements $s(i) \in S$ efficiently transfers all of $T$'s stored randomness to the digital detector. In practice, we probe $T$ with a set of mutually random binary SLM patterns $p(i) \in P$ to maximize photon throughput, and combine 2 separate layers of PDLC film into a thicker 1.5 mm stack to ensure $T$ is a fully random matrix.

Two issues prevent us from directly forming the set of speckle images $S$ into a collection of cryptographic keys. First, the histogram of an ideal speckle pattern's intensity values is exponentially distributed~\cite{Goodman:96}, whereas a key's must be uniformly distributed. We resolve this issue with digital whitening, which simply re-hashes detected bits together (via modulo addition) until they approach uniformity. We implement digital whitening by multiplying the detected speckle images (represented as a binary vector) with a large sparse binary matrix $H$~\cite{Zhou:11}. This technique is both computationally efficient and cryptographically secure: $H$ may be shared publicly without any loss to security (see Supplementary text for details).

\begin{figure}[b]
\centering
\includegraphics[scale=0.38]{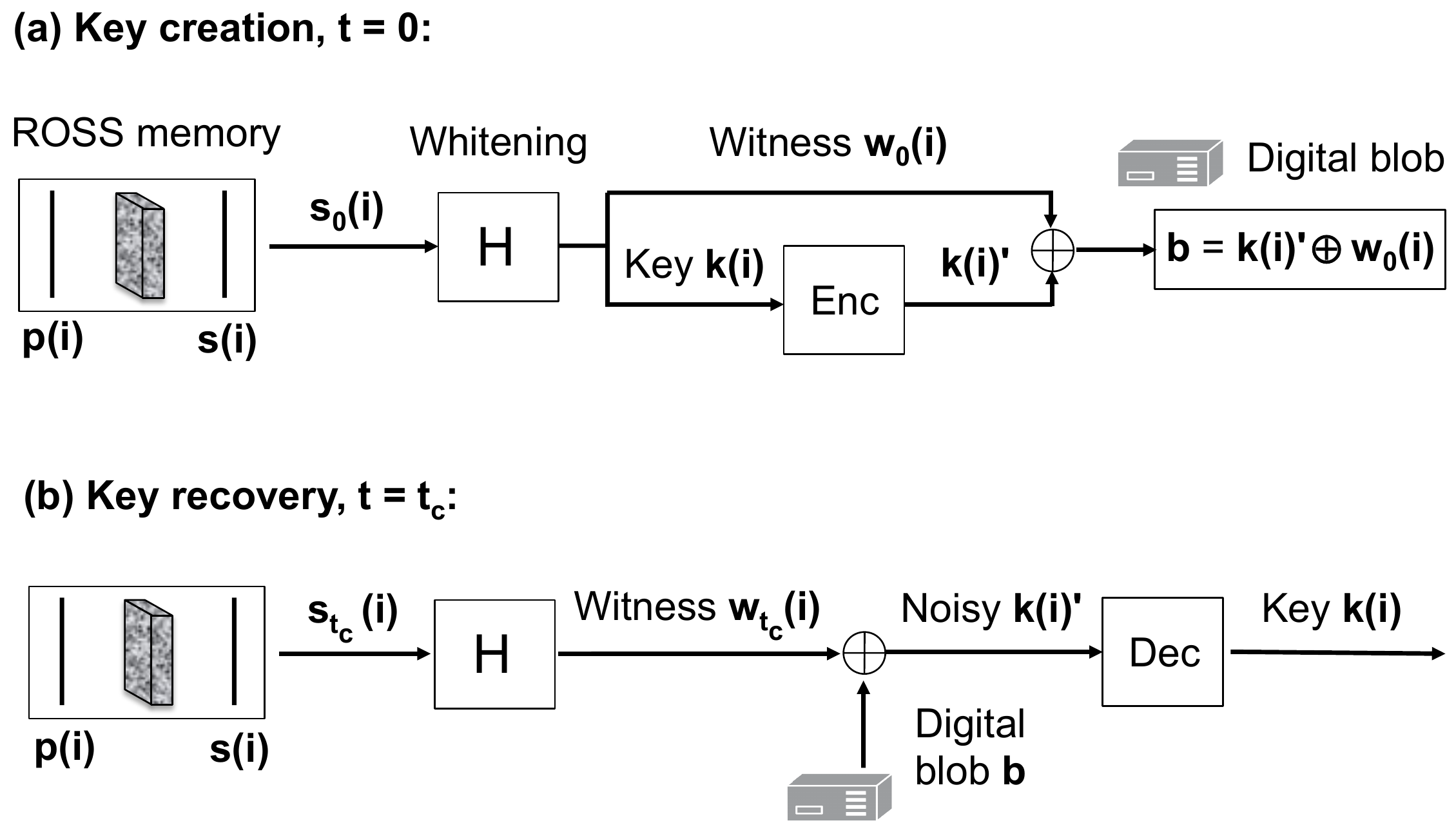}
\caption{The fuzzy commitment protocol. (a) A key $k$ and witness $w$ are created from the same whitened speckle image $s_0$, which are XOR'd together to create a secure, publicly sharable blob $b$. (b) The key is recovered at any later time $t_c$ by using the ROSS to re-create a noisy witness $w$, XORing it with $b$ and applying error correction to recover a noiseless key $k$.}
\label{fig4} 
\end{figure}

The second issue is the introduction of noise, which can prevent our device from perfectly recreating an initial random key $k(t=0)$ at a later time $t=t_c$. Absolute removal of all noise is a pre-requisite for effective non-digital key storage - a single flipped key bit will cause almost any cryptographic algorithm to crash. Noise is overcome by adopting a procedure called fuzzy commitment, an information-theoretically secure method of removing any flipped bits~\cite{Tuyls:07}. In short, fuzzy commitment sacrifices a fixed number of key bits to selectively remove any measurements that may have changed over time, using an error-correction algorithm. It consists of two steps: a ``creation" step performed once to generate a new key, and a ``recovery" step performed any subsequent time the same key is accessed, as detailed below. Additional details about fuzzy commitment are in the Supplementary text.

We experimentally test our ROSS device with fuzzy commitment by first displaying $n=4300$ random binary SLM screen inputs $p(i)$ to create and capture the same number of uncorrelated 4.85 MB speckle images $s(i)$. $n$ is  selected to efficiently extract all the randomness contained within our 1 mm$^3$ PDLC scattering volume without introducing unwanted correlations~\cite{Horstmeyer:13}. Each image $s(i)$ is transformed via the digital whitening matrix $H$ into a 2.42 MB vector. The same matrix $H$ transforms each image. We then implement key creation with fuzzy commitment, as outlined in Fig. 4(a). First, each whitened speckle sequence $s(i)$ is split into two segments: a 0.85 Mb key vector $k$ comprising a set of concatenated 332 random 256-bit encryption keys, and a 19.3 Mb witness vector $w$. Second, the key $k$ is encoded as a longer 19.3 Mb ``codeword" $k'$ and XOR'd with the witness to create an encrypted 19.3 Mb blob $b$, which we save digitally. The encrypted blob $b$ is information-theoretically secure, and may be shared publicly without any sacrifice to our system's physical security. 

At a later time $t_c=24$ hours, we attempt to access our saved keys using fuzzy commitment key recovery, as outlined in Fig. 4(b). First, we regenerate a noisy speckle image $s'(i)$. Second, we XOR $s'(i)$ with the saved blob $b$. Third, we apply a modified (255, 9) Hamming error correction with an additional 12.5\% data reduction factor to the XOR result, which recovers the key vector $k$ with minimal error. $k$ is subsequently split up into its 332 constituent secret keys for direct use. A single 256-bit key may be accessed by simply applying the above treatment to a cropped segment of one speckle image. 

\begin{figure}[t]
\centering
\includegraphics[scale=0.4]{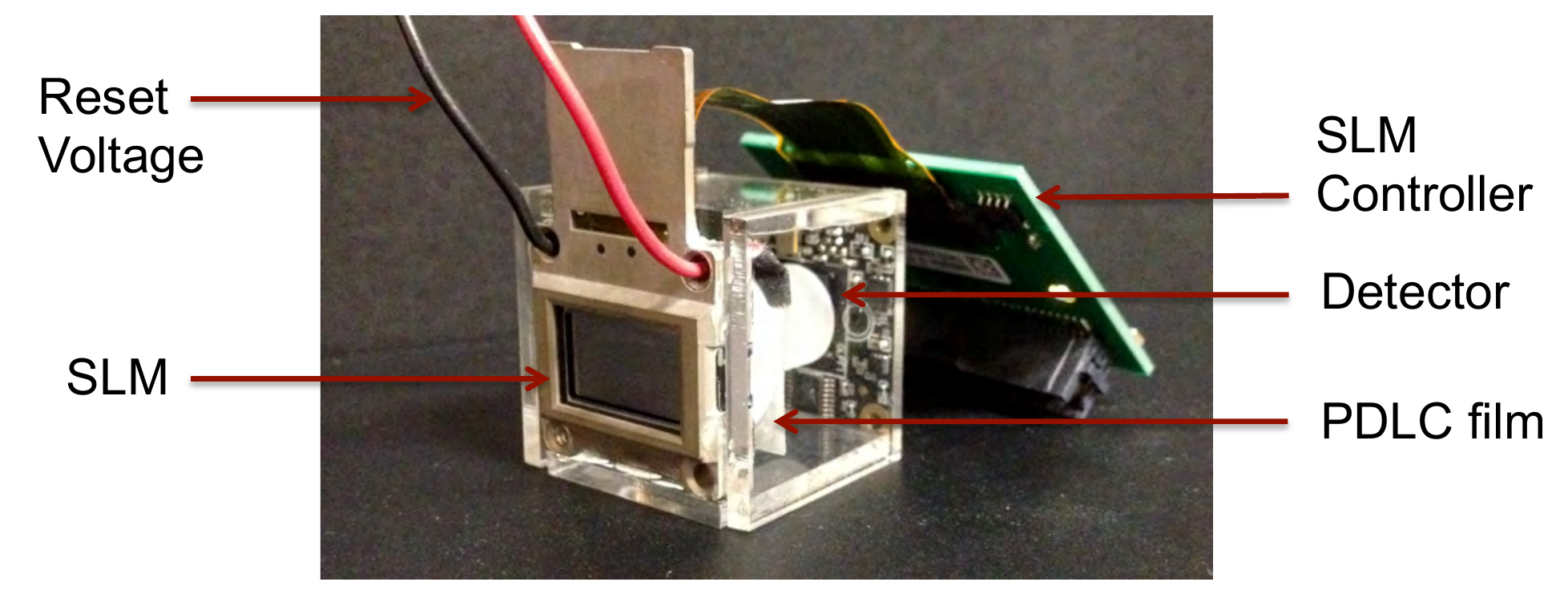}
\caption{An example experimental ROSS device. The optical source (not shown) illuminates the SLM from the left to form different speckle patterns on the detector.}
\label{fig5} 
\end{figure}

Executing key recovery with $n=4300$ unique speckle images at a later time $t_c$=24 hours leads to a total of 1.43 million 256-bit keys, of which 90.2\% are error-free. During practical operation, erroneous keys must be discarded and regenerated, which will delay the completion of any associated protocol. An example ROSS device is shown in Fig. 5. Fig.6(a) demonstrates our set of 256-bit ROSS keys are minimally correlated. A Gaussian fit of the key set's inter-key Hamming distance (i.e., key correlation) finds a mean of 0.50 and variance of $9.81 \times 10^{-4}$. Comparing this variance to the predicted variance of an independent, identically distribution binomial process ($9.77 \times 10^{-4}$) suggests each key comprises nearly 256 independent variables, as we expect. We additionally verify our ROSS key set's randomness by ensuring an arbitrarily selected 24 MB sequence of 750,000 concatenated keys passes all tests contained within the Diehard~\cite{Diehard} and NIST~\cite{NIST} statistical random number generator test suites. Example statistics from these tests are in the Supplementary text.

\begin{figure}[t]
\centering
\includegraphics[scale=0.27]{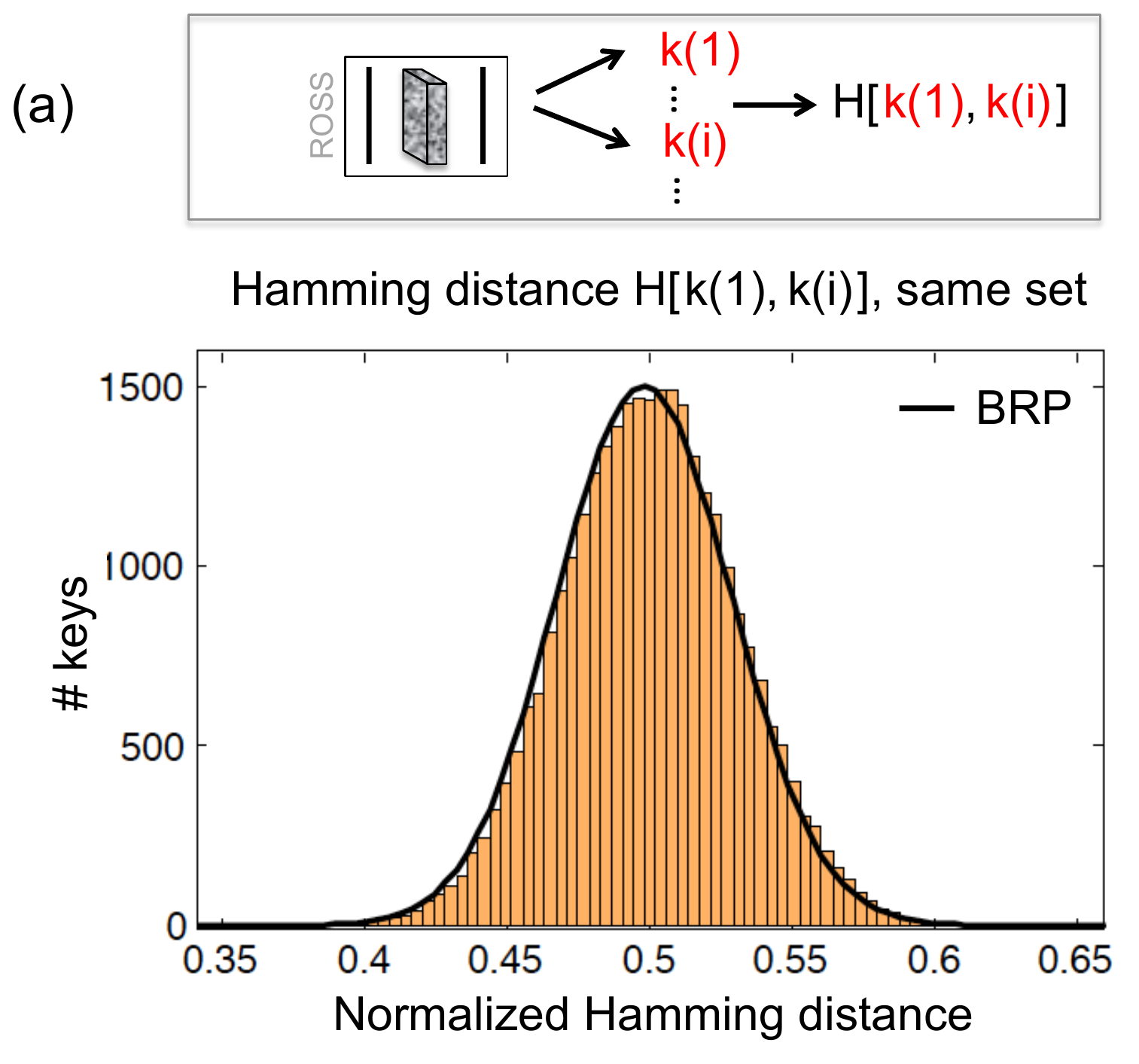}
\includegraphics[scale=0.27]{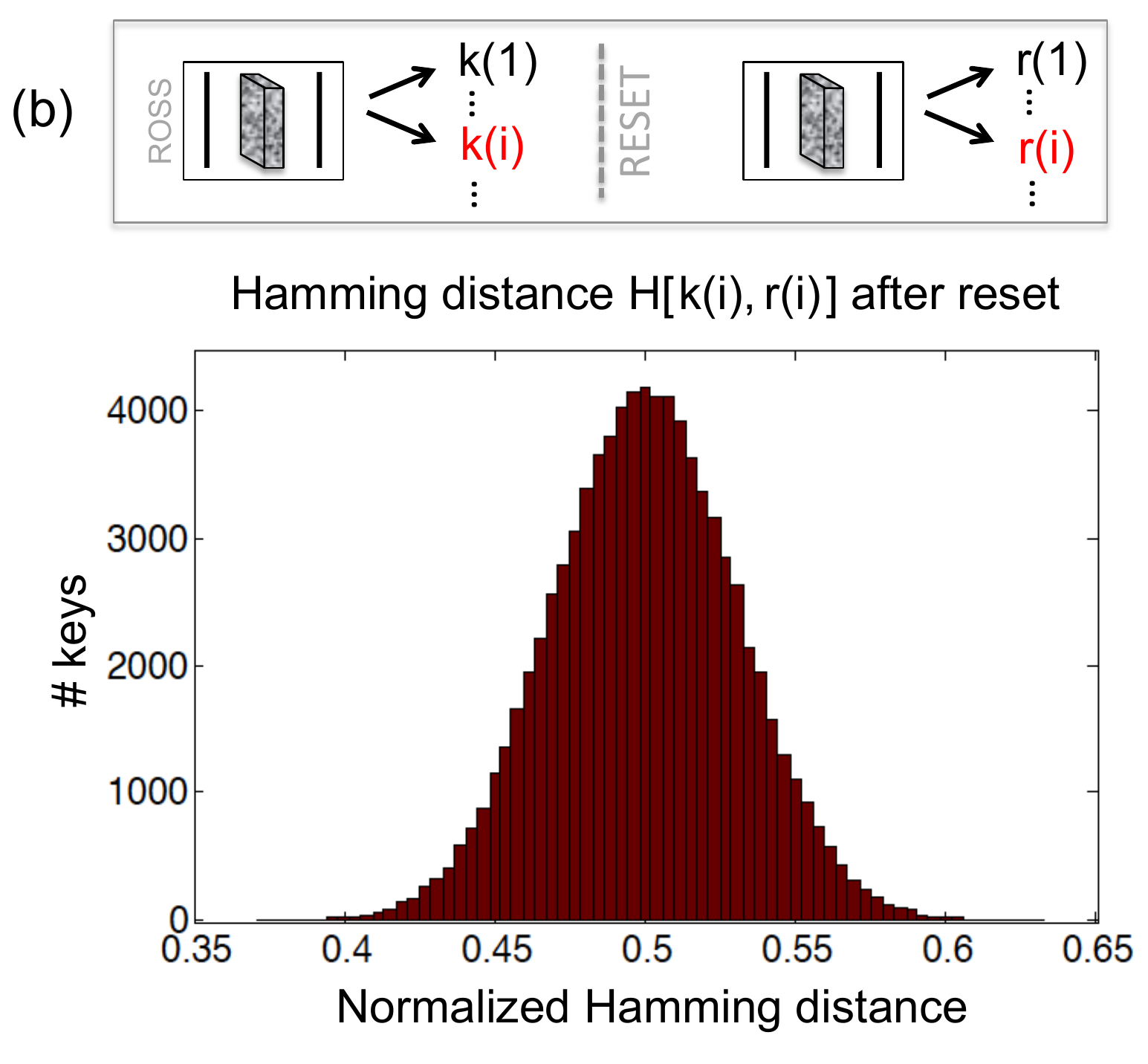}
\caption{(a) Hamming distance between one key $k(1)$ and $3 \times 10^{5}$ other keys from the same device follows an uncorrelated binary random process (BRP) curve. (b) Hamming distance between one key $k(i)$ and the corresponding key $r(i)$, creating using the same screen input $p(i)$ but after PDLC reset, similarly demonstrates complete key decorrelation. }
\label{fig5} 
\end{figure}

To demonstrate key reconfigurability, we use one ROSS device to generate 250 speckle images $S$ using the same SLM pattern set $P$ at four separate times: $t_1$-$t_4$. At $t_1$, we execute key creation to form key vector $k_1$ containing 8.4 x 10$^{4}$ individual 256-bit random keys. At $t_2$, two hours later, we perform key recovery to obtain 98\% of the keys in $k_1$ error-free. We then apply 40 V DC for 1 second across the PDLC interface to reset its scattering potential. At $t_3$, one minute after reset, we again display SLM pattern set $P$, but record a different set of speckle images, $S'$. Attempting to use $S'$ for key recovery of $k_1$ leads to the error histogram in Fig. 6(b). All random bits have been completely reset to a new uncorrelated configuration. However, we can use $S'$ to generate a new set of 8.4 x 10$^{4}$ keys, which we again recover two hours later at $t_4$ with 98\% accuracy. Our ROSS device thus continues to offer ``fresh" keys after each reset.

Currently, the primary shortcoming of our device is the limited lifetime over which a sequence of keys may be saved without error. The demonstrated 24 hour storage lifetime may potentially extend to several days or weeks, but decorrelation of the polymer film's optical response (due to temperature variation, movement, and laser source fluctuations) has been repeatedly observed. Material decorrelation also causes our high key error rate and limits the device's total number of saved keys by requiring low bit rate error correction. Future efforts will examine alternative optical setups that may better stabilize the scattering material, alternative error correction procedures that may help recover keys with higher bit rates, and different PDLC densities and film thicknesses with extended decorrelation half-lives.


To conclude, our reconfigurable optical PUF device is capable of storing over one million 256-bit keys within a physically disordered volumetric structure that is approximately 1 cubic millimeter in size. Keys may be reset into new, nearly perfectly uncorrelated sequences of random bits with a one-second applied DC voltage. To the best of our knowledge, no other device offers such a physically unclonable key storage medium that allows direct electronic reconfiguration. The potential density of stored randomness (10 Gb/mm$^3$) can offer many future cryptographic applications the opportunity to use multiple, large, time-varying keys to enhance security.   


\end{document}